\documentclass[12pt]{llncs2} 
\usepackage{ulem}
\usepackage{textcomp}
\usepackage{cite}
\usepackage{multicol}
\usepackage{amssymb}
\usepackage{amsmath}
\usepackage{comment}
\usepackage{graphicx}
\usepackage{mathrsfs}
\usepackage{booktabs,siunitx}
\usepackage{subcaption}
\usepackage{adjustbox}
\usepackage{multirow}
\usepackage{float} 
\usepackage{fancyhdr}
\usepackage{ragged2e}
\usepackage[%
square,        
comma,         
numbers,       
sort&compress, 
]{natbib}

\title{The Effect of Various Strengths of Noises and Data Augmentations on Classification of Short Single-Lead ECG Signals Using Deep Neural Networks}
\author{Faezeh Nejati Hatamian \and AmirAbbas Davari \and Andreas Maier}
\institute{%
	Pattern Recognition Lab, Department of Computer Science, Friedrich-Alexander-Universität Erlangen-Nürnberg, Erlangen, Germany\\%
}

\begin{document}
\emergencystretch 3em 
\maketitle



\section*{Background}

Due to the multiple imperfections during the signal acquisition, Electrocardiogram (ECG) datasets are typically contaminated with numerous types of noise, like salt and pepper and baseline drift. These datasets may contain different recordings with various types of noise~\cite{abdelazez2018detection} and thus, denoising may not be the easiest task. Furthermore, usually, the number of labeled bio-signals is very limited for a proper classification task.

\section*{Methods}

\begin{figure}[H]
	\begin{minipage}[b]{1\linewidth}
		\centering
		\centerline{\includegraphics[width=1\linewidth]{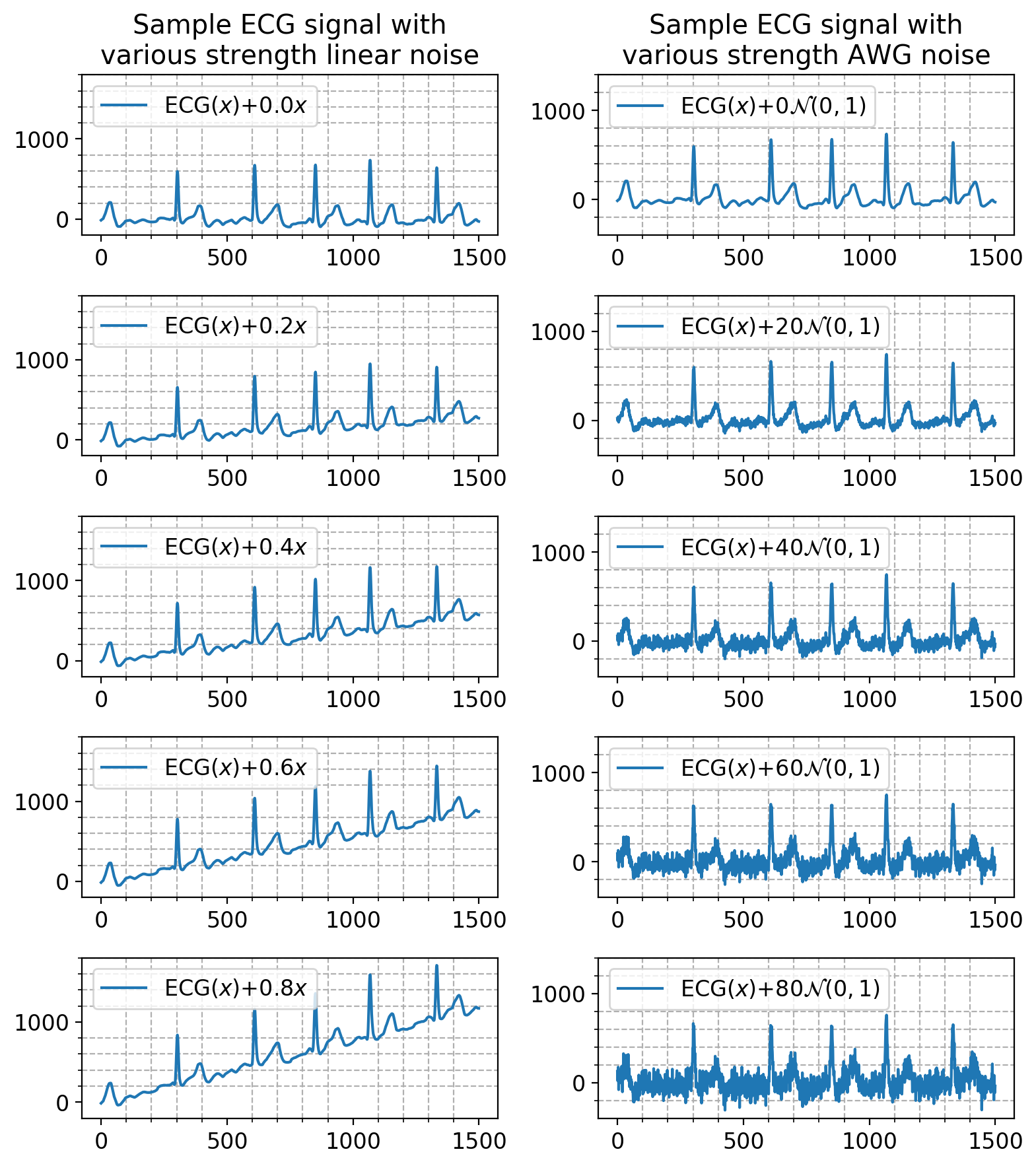}}
	\end{minipage}
	\caption{ Noisy ECG signal. Left: linear noise; right: AWGN.}
	\label{fig:sample_noisy_signals}
\end{figure}

In this work, we investigate the performance of deep convolutional neural networks and their separable alternatives (i.e., separable CNN) that are trained using noisy training data, tested on a clean test set. It is interesting to see if variants of CNNs and different schemes of data augmentation can help the classifier in the presence of noise contamination. To this end, we add different types and strengths of noise to the training data and train the network. We would like to see how noise in the training data impacts the inference performance and which noise is more destructive. For the purpose of simulation, we use additive white Gaussian noise (AWGN) in one variant and linear noise $\left(y = ax\right)$ in the other variant. In the case of the AWGN, we use 20, 40, 60, 80 values as the standard deviation to generate various strengths of this noise. Analogously, we use 0.2, 0.4, 0.6 and 0.8 as different slopes for the linear noise. Figure~\ref{fig:sample_noisy_signals} shows a sample signal from the training set being contaminated by different types and strengths of noise.

\begin{figure}[H]
	\begin{minipage}[c]{0.49\linewidth}
		\centering
		\centerline{\includegraphics[width=1\linewidth]{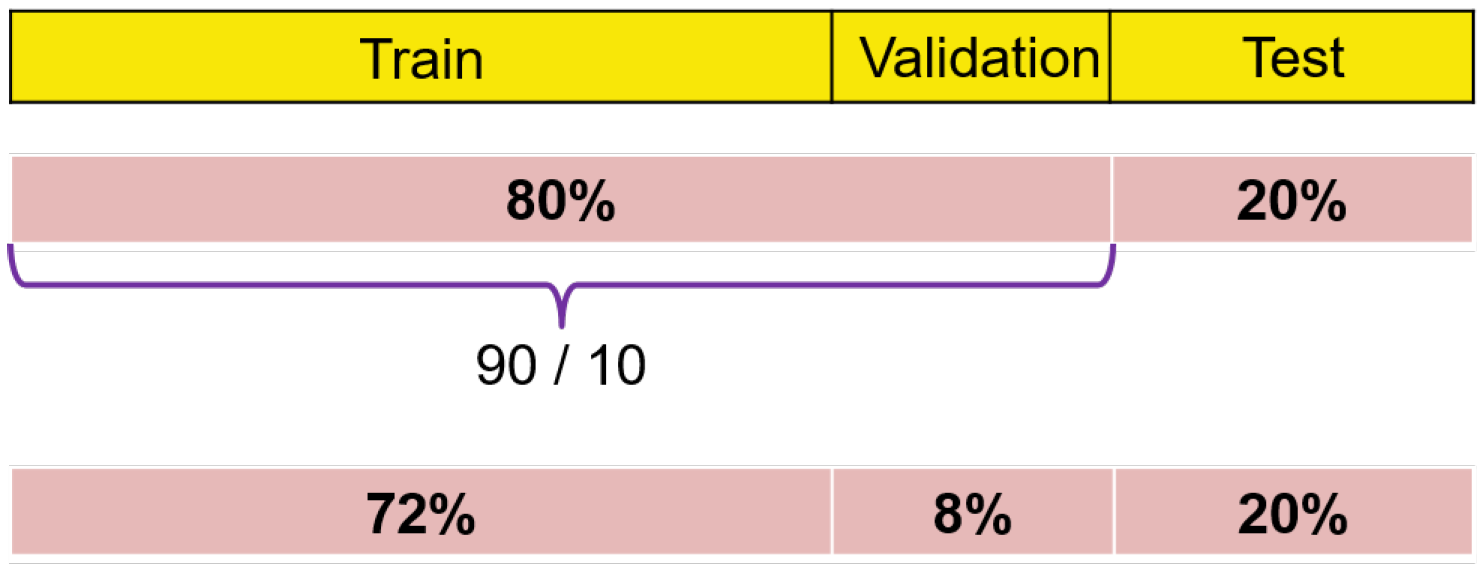}}
	\caption{dataset division policy.}
	\label{fig:dataset division policy.}
	\end{minipage}%
	\begin{minipage}[c]{0.49\linewidth}
		\centering
		\centerline{\includegraphics[width=1\linewidth]{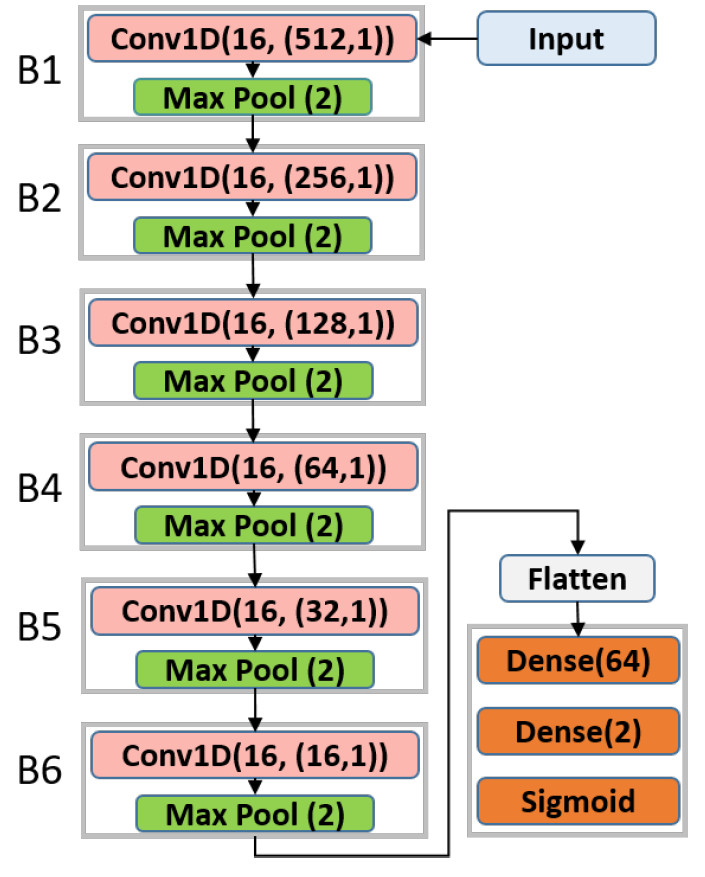}}
		\caption{CNN architecture.}
		\label{fig:CNN architecture.}
	\end{minipage}
\end{figure}

Moreover, we use oversampling and Gaussian Mixture Model (GMM)~\cite{davari2018gmm, davari2019fast} for data augmentation and compare the results with the variant without data augmentation. The metric that we adopt to measure the performance of the classifier is the F1-score. We use the same dataset~\cite{clifford2017af} in a similar fashion as in our previous work~\cite{hatamian2020effect} (Fig.~\ref{fig:dataset division policy.}). Finally, Fig.~\ref{fig:CNN architecture.} depicts the CNN architecture and its separable variant.

\section*{Results}

\begin{figure}[H]
	\begin{minipage}[b]{1\linewidth}
		\centering
		\centerline{\includegraphics[width=1\linewidth]{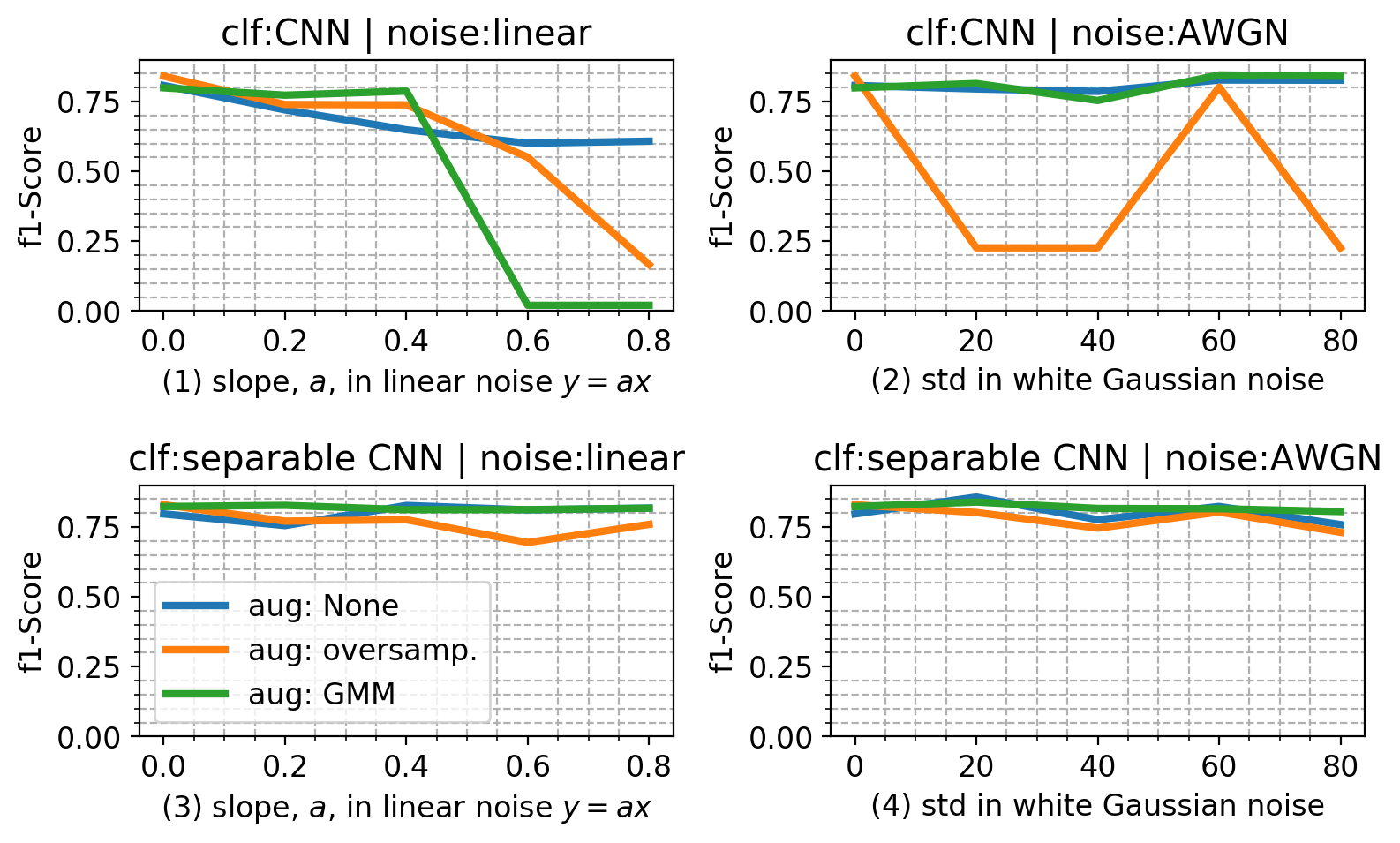}}
	\end{minipage}
	\caption{F1-scores vs. noise strengths for different noise (Linear and AWGN), data augmentation algorithms and classifiers.}
	\label{fig:F1_noise.}
\end{figure}

Figure~\ref{fig:F1_noise.} visualizes the F1-score vs. the corresponding noise strength for different data augmentation algorithms, types of noise and classifiers. Table~\ref{tab:F1-Classifier-Noise-Au} shows the detailed quantitative results.

\begin{table}[htbp]
	\centering
	\caption{F1-score for different classifiers and different noises (Linear, AWGN) with various strengths.}
	\resizebox{\textwidth}{!}{%
	\begin{tabular}{cc|c|c|c|c|c|c|c|c|c|c|}
		\cline{3-12}          &       & \multicolumn{5}{c|}{Slope in Linear Noise (a in y=ax)} & \multicolumn{5}{c|}{Standard Deviation in AWGN ($\sigma$ in $\mathcal{N}(0,\sigma))$} \\
		\hline
		\multicolumn{1}{|c|} {Classifier} & aug  & 0 & 0.2 & 0.4 & 0.6 & 0.8   & 0 & 20 & 40 & 60 & 80 \\
		\hline
		\multicolumn{1}{|c|}{\multirow{2}{*}{CNN}} & None  & 80.76 & 72.02 & 64.89 & 60.08 & 60.81 & 80.76 & 79.50 & 78.63 & 82.71 & 82.63 \\
		\multicolumn{1}{|c}{} & GMM   & 79.91 & 77.29 & 78.74 & 02.06 & 02.08  & 79.91 & 81.45 & 75.38 & 84.53 & 84.07 \\
		\hline
		\multicolumn{1}{|c|}{\multirow{2}{*}{Separable CNN}} & None  & 79.69 & 75.54 & 82.72 & 81.15 & 81.79 & 79.69 & 85.73 & 77.62 & 82.37 & 75.89 \\
		\multicolumn{1}{|c}{} & GMM   & 82.33 & 82.76 & 81.22 & 81.24 & 81.75 & 82.33 & 83.95 & 81.65 & 81.56 & 80.55 \\
		\hline
	\end{tabular}}%
	\label{tab:F1-Classifier-Noise-Au}%
\end{table}%

\section*{Conclusion}

It can be observed that in the presence of various types and strengths of the noise, the separable CNN is more stable and robust than the normal CNN. In the majority of the experiments, linear noise is more destructive than the AWGN. In fact, both CNN and separable CNN perform equally well in the presence of the AWGN. However, CNN simply falls apart in the presence of the linear noise. In this case, the separable CNN brings robustness to the classification pipeline. Finally, except in the case of linear noise and normal CNN, the highest classification performance is almost consistently being achieved when the GMM data augmentation is in use.


\renewcommand{\bibsection}{\section*{References}} 
\bibliographystyle{splncsnat}
\begingroup
\small 
\bibliography{References}
\endgroup

\ \\
%

\end{document}